\documentclass[aps,prb,twocolumn,superscriptaddress,amsmath,amssymb,floatfix]{revtex4-2}

\usepackage{amsmath,amssymb,amsfonts}      
\usepackage{graphicx}                      
\usepackage{dcolumn}                      
\usepackage{bm}                           
\usepackage{hyperref}                     
\usepackage[usenames,dvipsnames]{xcolor}  
\usepackage{booktabs}                     
\usepackage{siunitx}                      
\usepackage{mhchem}                  
\usepackage[T1]{fontenc}                  
\usepackage{rotating}
\usepackage[toc,page]{appendix}
\usepackage{multirow}

\usepackage{placeins} 

\hypersetup{
  colorlinks=true,
  linkcolor=BlueViolet,
  citecolor=ForestGreen,
  urlcolor=MidnightBlue
}

\begin{document}

\title{Crystal-Field--Driven Magnetoelectric Coupling in the Non-Kramers Hexaaluminate PrMgAl$_{11}$O$_{19}$}

\author{Sonu Kumar}
\email{sonu.kumar@matfyz.cuni.cz}
\affiliation{Charles University, Faculty of Mathematics and Physics, Department of Condensed Matter Physics, Prague, Czech Republic}
\affiliation{Adam Mickiewicz University, Faculty of Physics and Astronomy, Department of Experimental Physics of Condensed Phase, Pozna\'n, Poland}

\author{Ga\"{e}l Bastien}
\affiliation{Charles University, Faculty of Mathematics and Physics, Department of Condensed Matter Physics, Prague, Czech Republic}

\author{Ross H.\ Colman}
\affiliation{Department of Condensed Matter Physics, Faculty of Mathematics and Physics, Charles University, Prague, Czech Republic}

\author{Maxim Savinov}
\affiliation{Institute of Physics, Czech Academy of Sciences, Prague, Czech Republic}

\author{Petr Proschek}
\affiliation{Charles University, Faculty of Mathematics and Physics, Department of Condensed Matter Physics, Prague, Czech Republic}

\author{Michal Vali\v{s}ka}
\affiliation{Charles University, Faculty of Mathematics and Physics, Department of Condensed Matter Physics, Prague, Czech Republic}

\author{Mateusz Kempiński}
\affiliation{Adam Mickiewicz University, Faculty of Physics and Astronomy, Department of Experimental Physics of Condensed Phase, Poznań, Poland}

\author{Wojciech Kempiński}
\affiliation{Institute of Molecular Physics, Polish Academy of Sciences,
Department of Low Temperature Physics, Quantum Materials and Technologies, Poznań, Poland}

\author{Ma\l{}gorzata \u{S}liwi\'nska-Bartkowiak}
\affiliation{Adam Mickiewicz University, Faculty of Physics and Astronomy, Department of Experimental Physics of Condensed Phase, Pozna\'n, Poland}

\author{Stanislav Kamba}
\email{kamba@fzu.cz}
\affiliation{Institute of Physics, Czech Academy of Sciences, Prague, Czech Republic}

\date{\today}

\begin{abstract}
We report broadband dielectric spectra of the non-Kramers hexaaluminate
PrMgAl\textsubscript{11}O\textsubscript{19}, revealing a pronounced interplay between
permittivity and magnetization at cryogenic temperatures. The zero-field dielectric
response follows a Barrett-type quantum-paraelectric form, while a broad dielectric
anomaly near \SI{5}{\kelvin} shows a complex field dependence that mirrors the multi-hump
behavior of the magnetic specific heat, evidencing robust magnetoelectric coupling.
The inverse permittivity $\varepsilon'^{-1}(T,H)$ scales linearly with $M^2$, consistent
with a biquadratic $P^2M^2$ term in a Landau framework.
Fits yield a temperature-dependent coupling constant $\lambda(T)$ that decreases with
heating from $(1.07\pm0.01)\times10^{-4}\,\mu_{\mathrm{B}}^{-2}$ (at \SI{5}{\kelvin}) to
$(4.77\pm0.02)\times10^{-5}\,\mu_{\mathrm{B}}^{-2}$ (at \SI{10}{\kelvin}), reflecting the thermal population of low-lying
energy levels of Pr$^{3+}$. Consistently, the uniaxial thermal expansion develops an additional low-temperature
hump below $\sim\SI{30}{\kelvin}$ that is progressively suppressed by magnetic field, recovering an approximately
saturated response by \SI{9}{\tesla}.
These results identify PrMgAl\textsubscript{11}O\textsubscript{19} as a paradigmatic
non-Kramers hexaaluminate where quantum paraelectricity and magnetoelectric interactions
are intrinsically entangled, establishing hexaaluminates as a tunable platform for
magnetoelectric physics in frustrated quantum materials.
\end{abstract}

\maketitle

\section{Introduction}

Magnetoelectric (ME) coupling, the interconversion of electric and magnetic fields, was first proposed by Pierre Curie in the 19th century and experimentally realized in 1960 by Astrov in \ce{Cr2O3} \cite{Astrov1960}. This discovery initiated the study of materials with coupled ferroic orders. While linear ME effects rely on the $P_iM_j$ invariant, nonlinear responses dominate when time-reversal or inversion symmetry forbids it \cite{Shvartsman2010}. Microscopically, ME effects arise from mechanisms like symmetric exchange striction in collinear magnets \cite{Sergienko2006}, the inverse Dzyaloshinskii--Moriya interaction in non-collinear systems \cite{Katsura2005,Sergienko2006}, and spin-dependent metal--ligand hybridization \cite{Arima2007,Jia2007}. Initially observed in antiferromagnetic \ce{Cr2O3} \cite{Astrov1960}, ME coupling studies extend to magnetoelectric multiferroics, where the highest ME coupling is observed \cite{Eerenstein2006,dong2015multiferroic}, but also to quantum magnets like \ce{TlCuCl3} \cite{Kimura2017}.

Frustrated systems introduce new ME research avenues, where competing interactions stabilize exotic ground states. In magnetoplumbites, off-centered M$^{3+}$ ions (M = Al, Ga, Fe) in \ce{MO5} trigonal bipyramids form electric dipoles on a triangular lattice, creating an Ising-like ``frustrated-dipole'' system \cite{Albanese1992,Iyi1990,Kimura1990,Cao2015,Holtstam2020,Rowley2016,Rensen1969,Shen2014}. \ce{EuAl12O19} exemplifies this, hosting a dynamically disordered ``antipolar liquid'' with short-range polar correlations but no long-range (anti)ferroelectric order \cite{bastien2024frustrated}, establishing hexaaluminates as a platform for dielectric liquids. The M$^{3+}$ ion’s off-centering in \ce{MO5} bipyramids, driven by a shallow double-well potential and soft phonons, produces Ising dipoles \cite{Wang2014PRX_AFE}. Dipole--dipole interactions favor antipolar in-plane and polar interplane correlations, forming a frustrated antiferroelectric \cite{Li2016_ZNBS,Kimura1990,Rowley2016,Cao2015,Zhang2020,Zhang2024,Shen2016}. In \ce{EuAl12O19}, dielectric spectroscopy shows a relaxation mode softening from THz to Hz upon cooling, yielding high permittivity, with a second-order transition near \SI{49}{\kelvin} enhancing the response but maintaining a liquid-like antipolar state \cite{bastien2024frustrated}.

Rare-earth hexaaluminates with the space group $P6_{3}/mmc$ provide a platform where frustrated \ce{AlO5} dipoles couple to $4f$ magnetism. In \ce{CeMgAl11O19}, the Kramers-protected ground doublet couples to dipoles through spin--orbit entanglement, giving rise to a predominantly biquadratic $P^{2}M^{2}$ magnetoelectric (ME) coupling~\cite{Kumar_CeMgAl11O19}. In contrast, \ce{PrMgAl11O19} hosts a non-Kramers Pr$^{3+}$ quasi-doublet that is split by local symmetry breaking~\cite{Kumar2025_PRB,Cao2024_SynthesisDisorderIsing_PrMgAl11O19}. Notably, local symmetry lowering has been observed experimentally across the isostructural hexaaluminate family\cite{Kumar2025_NdMgAl11O19,Bastien2025_CeMgAl,Kumar2026_SmMgAl11O19}. Rather than producing a trivial nonmagnetic ground state, the quasi-doublet splitting acts as an intrinsic transverse field which, together with exchange interactions, gives rise to induced quantum magnetism at low temperatures. However, at finite temperatures, antiferromagnetic interactions promote magnetic moments through the population of the excited singlet, a phenomenon referred to as induced quantum magnetism \cite{Thalmeier2024_CEF}.

In this article, we investigate the magnetoelectric coupling of \ce{PrMgAl11O19} through combined dielectric permittivity and magnetization measurements under applied magnetic fields.
The objectives of this study are threefold: (i) to establish the low-temperature dielectric response of \ce{PrMgAl11O19} under magnetic field and quantify the field evolution of the dielectric anomaly, (ii) to test whether the magnetodielectric response is governed by a Landau-type biquadratic coupling through the scaling $\varepsilon'^{-1}(T,H)\propto M^{2}$, and (iii) to extract an effective temperature-dependent coupling strength $\lambda(T)$ and compare it to Kramers hexaaluminate analogues.
The limitations of the present work are that the dielectric anomalies are broad and partially overlapping, and the magnetic response can contain multiple contributions from low-lying crystal electric field (CEF) levels, which complicates a unique separation of possible cubic ($PM^{2}$) and biquadratic ($P^{2}M^{2}$) terms. Consequently, the Landau analysis should be viewed as a quantitative phenomenological description of the coupled response.

In the isostructural compound \ce{EuAl12O19}, the magnetic and dielectric responses are essentially independent, with no measurable magnetoelectric coupling reported \cite{bastien2024frustrated}. By contrast, in \ce{PrMgAl11O19},
dielectric measurements reveal a Barrett-like quantum-paraelectric response with a broad low-temperature anomaly, while $\varepsilon'^{-1}(T,H)$ scales linearly with $M^{2}$, consistent with a quadratic channel and enabling a quantitative extraction of the biquadratic ME coupling strength and its temperature evolution $\lambda(T)$.
These findings highlight \ce{PrMgAl11O19} as a non-Kramers hexaaluminate in which quantum paraelectricity, magnetic frustration, and magnetoelectricity coexist, providing a route to entangled ME states beyond the Kramers paradigm.

\section{Experimental details}

The synthesis and single-crystal growth of \ce{PrMgAl11O19} were carried out using a combination of solid-state reaction and the optical floating-zone technique. High-purity precursor oxides (\ce{Pr6O11}, MgO, and \ce{Al2O3}; 99.99\% purity, Sigma Aldrich) were first calcined at \SI{800}{\celsius} for 24~h in air to remove moisture and carbonate contamination. After calcination, the powders were weighed in stoichiometric proportion, thoroughly mixed, and ground in an agate mortar to ensure homogeneity. The mixture was pressed into cylindrical rods of \SI{6}{\milli\metre} diameter and \SI{100}{\milli\metre} length under a quasihydrostatic pressure of 2~tons for 15~min. The resulting rods were sintered at \SI{1200}{\celsius} for 72~h in air to complete the solid-state reaction and to densify the material.

Crystal growth was performed using a four-mirror optical floating-zone furnace (Crystal Systems Corp.) under a flowing air atmosphere with a slight overpressure of 1~atm to suppress Pr volatility. The airflow rate was maintained at \SI{3}{\litre\per\minute}. Both feed and seed rods were taken from the sintered material. During growth, the rods were counter-rotated at 30~rpm to homogenize the molten zone, and the growth rate was fixed at \SI{2}{\milli\metre\per\hour}. Further details of the synthesis and growth procedure are reported in Refs.~\cite{Kumar2025_PRB,Kumar2025_NdMgAl11O19,Bastien2025_CeMgAl}. The resulting ingot was dark green in colour and contained multiple large grains separated by visible grain boundaries. Single grains were isolated using a wire saw and mechanical cleavage. The single-crystalline nature of each grain was verified by back-reflection Laue x-ray diffraction.

For dielectric measurements, selected single-crystal grains free of visible twins were oriented and cut perpendicular to the crystallographic $c$ axis, then lapped to optical flatness. The opposing faces were sputtered with an \SI{80}{\nano\metre} Au layer to form parallel-plate electrodes. Complex permittivity in the frequency range \SI{10}{\hertz}--\SI{1}{\mega\hertz} was measured using a Novocontrol Alpha-A impedance analyser mounted in a $^3$He cryostat (\SIrange{0.3}{300}{\kelvin}) equipped with a \SI{9}{\tesla} superconducting solenoid. At each temperature point, the crystal was zero-field cooled; both electric and magnetic fields were applied parallel to the $c$ axis.

Magnetization was measured using a Quantum Design MPMS7 SQUID magnetometer on a single crystal (mass = 4.38~mg) from the same growth batch as the dielectric sample. To confirm reproducibility, additional measurements were performed on a different crystal (mass = 2.6~mg) from a separate growth batch. The results were consistent with those reported previously in Ref.~\cite{Kumar2025_PRB}, confirming the intrinsic magnetic behaviour of \ce{PrMgAl11O19}.

Thermal expansion measurements were performed using a high-resolution capacitive dilatometer installed in a Quantum Design PPMS-9 system. The relative length change was recorded as a function of temperature in both zero magnetic field and under applied magnetic fields. Measurements were carried out for two crystallographic orientations, i.e. in the \textit{ab} plane and along the $c$ axis.

Electron paramagnetic resonance (EPR) spectra were recorded using a Radiopan ES/X spectrometer equipped with an Oxford Instruments helium-flow cryostat, covering the temperature range \SIrange{1.5}{300}{\kelvin}. Single-crystal fragments were mounted on a rotatable quartz holder to probe the anisotropy of the EPR signal.

\section{Results}

\begin{figure}
\centering

\includegraphics[width=0.48\textwidth]{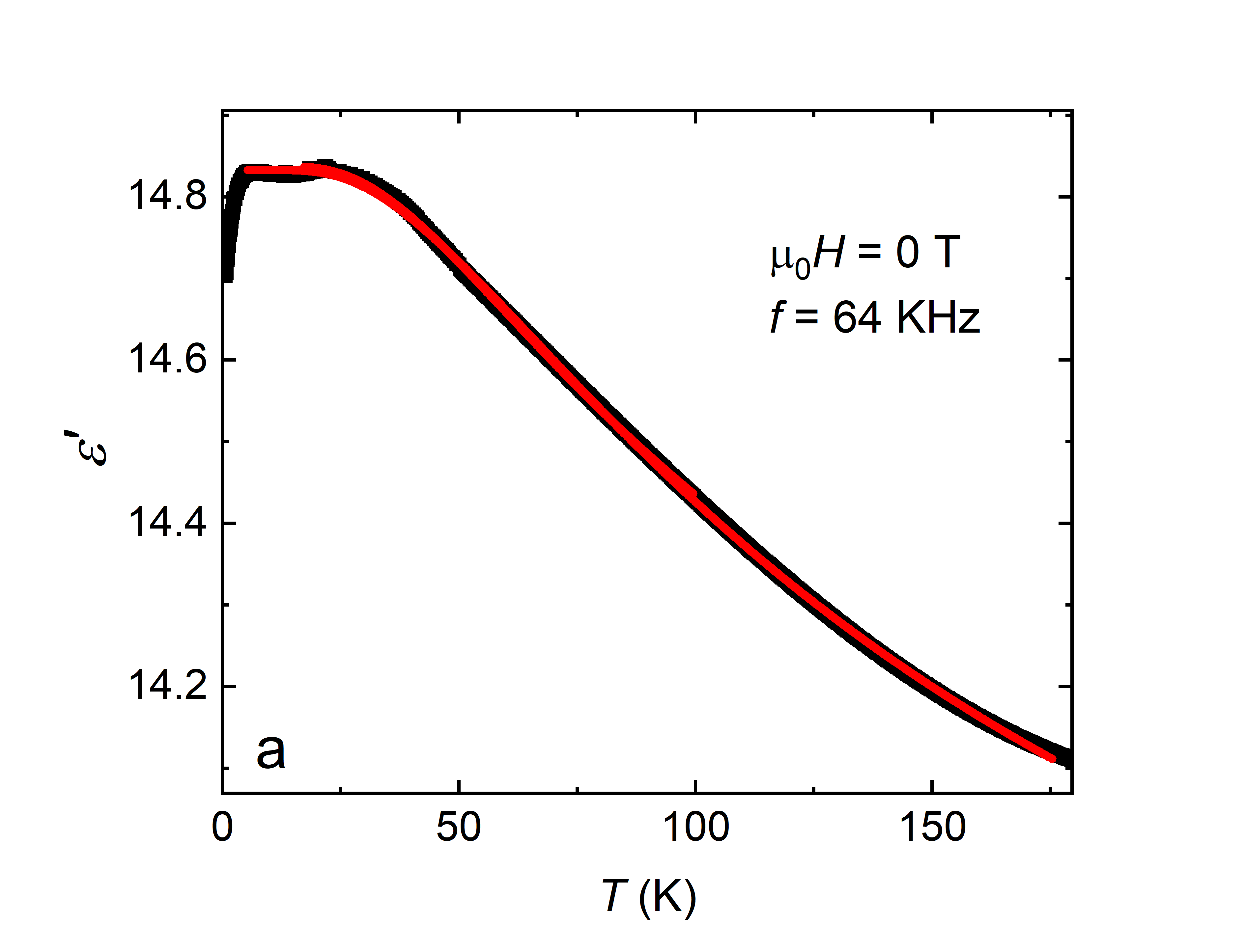}\hfill
\includegraphics[width=0.48\textwidth]{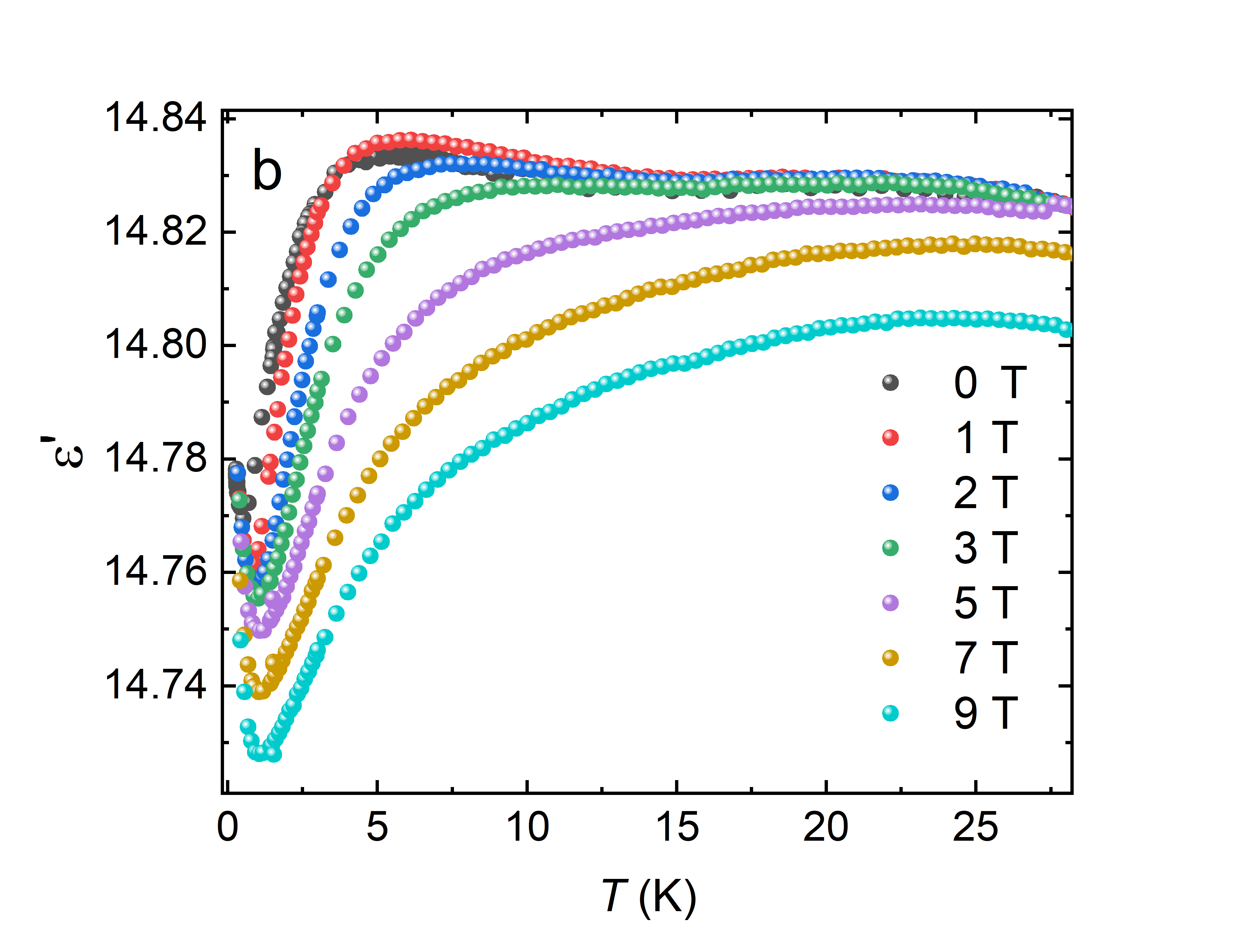}

\includegraphics[width=0.48\textwidth]{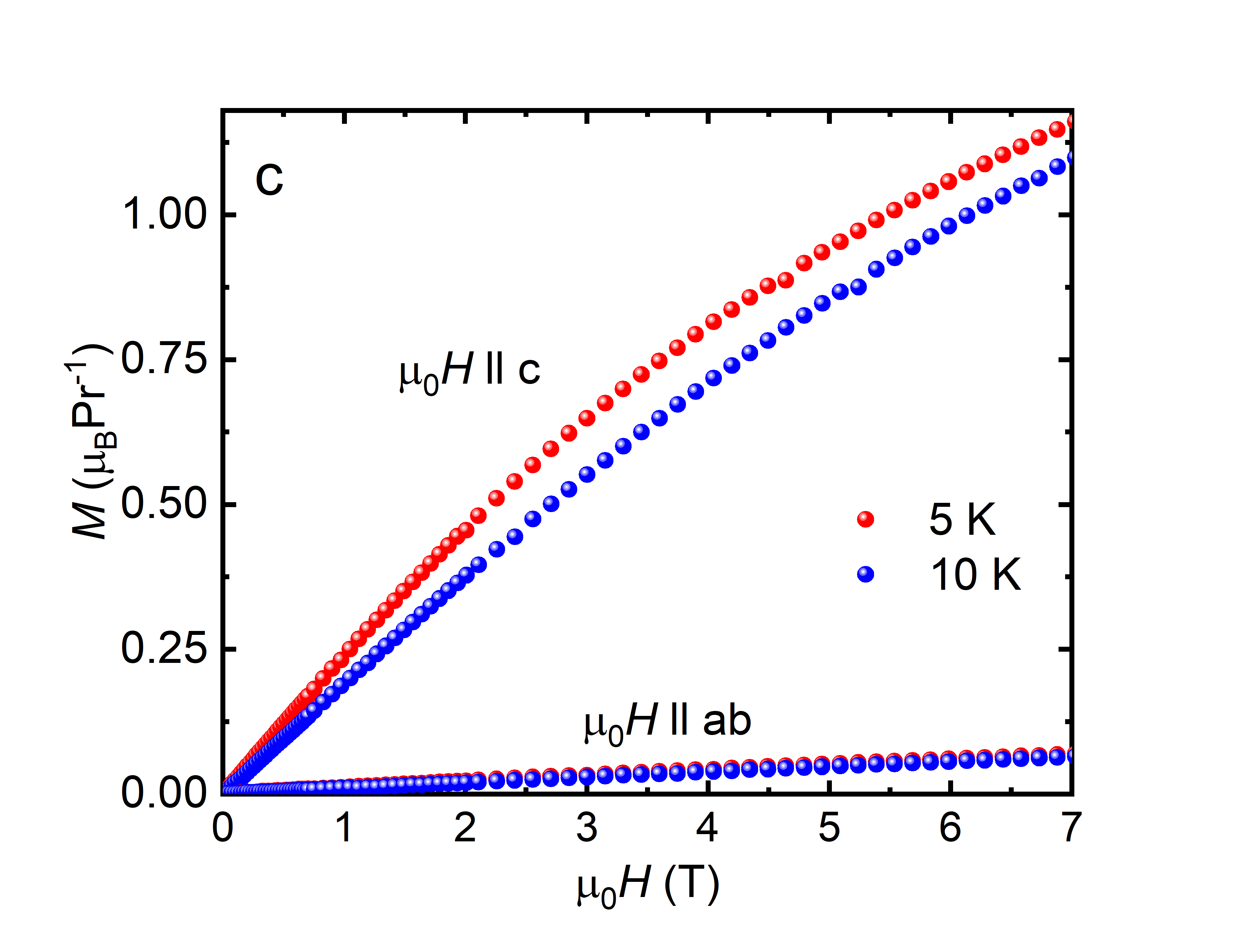}

\caption{(a) Zero-field dielectric permittivity $\varepsilon'(T)$ with a Barrett fit (solid red line) from 10--170\,K.
(b) Temperature-dependent dielectric permittivity $\varepsilon'(T,H)$ measured at 64\,kHz in different magnetic fields.
(c) Isothermal magnetization $M(H)$ at 5 and 10\,K for fields applied along the $c$ axis and in the $ab$ plane, highlighting strong Ising anisotropy.}
\label{fig:Fig1}
\end{figure}

The zero-field permittivity $\varepsilon'(T)$ of \ce{PrMgAl11O19} increases smoothly with decreasing temperature and begins to saturate below \SI{30}{\kelvin}, consistent with the behaviour of an incipient quantum paraelectric (Fig.~\ref{fig:Fig1}a).

A fit to the Barrett expression \cite{Barrett1952,Rowley2014},
\begin{equation}
\varepsilon'(T) = \varepsilon_{\infty} + \frac{M}{\tfrac{1}{2}T_{1}\coth\!\left(\tfrac{T_{1}}{2T}\right) - T_{0}},
\end{equation}
describes the data accurately down to \SI{20}{\kelvin}. The fit yields
$\varepsilon_{\infty} = 13.21 \pm 0.01$,
$M = 244.7 \pm 4.7$~K,
$T_{0} = -88.3 \pm 2.3$~K,
and $T_{1} = 124.6 \pm 0.7$~K.
In this framework, $\varepsilon_{\infty}$ represents the background permittivity from high-energy excitations (electrons and phonons), $M$ reflects the dielectric strength of the soft polar mode, $T_{1}$ sets the energy scale of quantum fluctuations, and $T_{0}$ is the extrapolated Curie temperature. The strongly negative $T_{0}$ demonstrates that quantum fluctuations suppress a ferroelectric transition, confirming that \ce{PrMgAl11O19} is a quantum paraelectric in zero field.

Below $\sim\SI{20}{\kelvin}$ the permittivity starts to deviate from the Barrett prediction and develops a broad low-$T$ anomaly centered near $\sim\SI{5}{\kelvin}$.
Its evolution with magnetic field is clearly field dependent, but not strictly monotonic: at low fields the feature is strongly broadened and partially overlaps between $\mu_0H=0$--1~T, while for higher fields it progressively shifts and merges into the higher-temperature shoulder (Fig.~\ref{fig:Fig1}b).
This complex field response closely mirrors the behavior of the magnetic specific heat observed in the sample \cite{Kumar2025_PRB}, where the low-$T$ hump is not purely Schottky-like below $\sim\SI{4}{\tesla}$ and shows pronounced overlap in the 0--1~T range.
In addition, a second broad hump is visible in $\varepsilon'(T)$ near $\sim\SI{30}{\kelvin}$, which remains essentially field insensitive, in direct correspondence with the higher-temperature structure observed in $C_m/T$.
Since the double-hump structure in $C_m/T$ is attributed to crystal-field effects, the fact that the permittivity reproduces the same two-scale phenomenology indicates that the dielectric response is directly coupled to the same low-energy CEF excitations of Pr$^{3+}$.

Magnetization measurements with fields applied parallel and perpendicular to the crystallographic $c$ axis (Fig.~\ref{fig:Fig1}c) confirm pronounced Ising anisotropy, consistent with previous results~\cite{Kumar2025_PRB} and EPR measurements (see the Supplementary Information) \cite{SI,AbragamBleaney1986,Bodziony2008JAC,Weil2006}. For $H \parallel c$, $M(H)$ at 5--10~K increases nonlinearly up to \SI{7}{\tesla}, reaching $\sim$1.16~$\mu_{\mathrm{B}}$/Pr, while for $H \perp c$ it remains linear and much smaller ($\sim$0.05~$\mu_{\mathrm{B}}$/Pr). The agreement with earlier measurements~\cite{Kumar2025_PRB} confirms reproducibility, and these data are remeasured here for quantitative comparison with the dielectric response.

\section{Discussion}
\subsection{Quantum-paraelectric background and field evolution of the permittivity}

To understand the observed magnetoelectric coupling in \ce{PrMgAl11O19}, we begin with a phenomenological Landau free-energy functional that incorporates both electric polarization $P$ and magnetization $M$~\cite{Harris2008}:
\begin{align}
F &= \tfrac{1}{2} \alpha_P(T) P^2 + \tfrac{1}{4} b_P P^4
     + \tfrac{1}{2} \alpha_M(T) M^2  \nonumber\\
  &\quad + \tfrac{1}{4} b_M M^4
     + \eta P M^2 + \lambda P^2 M^2 .
\label{eq:landau}
\end{align}
where $\alpha_P(T)$ and $\alpha_M(T)$ are the dielectric and magnetic stiffnesses, $b_P$ and $b_M$ are higher-order anharmonic coefficients, $\lambda$ quantifies the biquadratic magnetoelectric coupling (the leading term allowed by the global $P6_3/mmc$ inversion symmetry), and $\eta$ denotes an effective cubic coupling that can arise from local distortions.

Minimizing with respect to $P$ in the small-$P$ limit gives
\begin{equation}
\frac{\partial F}{\partial P} = \alpha_P(T) P + \eta M^2 + 2 \lambda P M^2 = 0,
\end{equation}
so that
\[
P = -\frac{\eta M^2}{\alpha_P(T) + 2 \lambda M^2}.
\]
For $\lambda M^2 \ll \alpha_P(T)$, this reduces to $P \approx -\eta M^2 / \alpha_P(T)$. Including an external electric field via $-P E$, the dielectric susceptibility is
\[
\chi_e = \frac{1}{\alpha_P(T) + 2 \lambda M^2},
\]
which implies
\begin{equation}
\varepsilon'^{-1}(T,H) \approx \alpha_P(T) + 2 \lambda M^2(T,H).
\end{equation}

Thus $\varepsilon'^{-1}$ should scale linearly with $M^2$. Fits at \SI{5}{\kelvin} and \SI{10}{\kelvin} confirm this (Fig.~\ref{fig:Fig3}(a) and \ref{fig:Fig3}(b)), yielding
\begin{align}
\lambda_{5\,\mathrm{K}}  &= (1.07 \pm 0.01)\times 10^{-4}\,\mu_B^{-2}, \nonumber\\
\lambda_{10\,\mathrm{K}} &= (4.77 \pm 0.02)\times 10^{-5}\,\mu_B^{-2}.
\end{align}
with $\alpha_P(T) \approx 0.0674$. The reduction of $\lambda$ with increasing $T$ signals weaker magnetoelectric coupling with the population of the excited singlet.

A second hallmark of magnetoelectricity is the low-temperature permittivity anomaly.
In \ce{CeMgAl11O19} this appears as a broad dielectric dip, whose anomaly temperature follows a clear magnetization scaling and shifts approximately quadratically with $M^{2}$ upon applying magnetic field~\cite{Kumar_CeMgAl11O19}.
Notably, the zero-field anomaly occurs near \SI{3.6}{\kelvin}, i.e., in a predominantly paramagnetic regime where the magnetization curves are well captured by a Brillouin-function description of essentially non-interacting ions~\cite{Kumar_CeMgAl11O19}.
By contrast, in \ce{PrMgAl11O19} the dielectric response exhibits a broad hump near \SI{5}{\kelvin} (Fig.~\ref{fig:Fig1}b), whose field evolution is not governed by a single field-tunable Schottky gap.
Instead, the low-$T$ anomaly develops within a regime of strong exchange renormalization~\cite{Kumar2025_PRB} and its field dependence closely mirrors the complex multi-hump structure observed in $C_{m}/T$.
In addition, \ce{PrMgAl11O19} shows a second, field-insensitive hump around $\sim\SI{30}{\kelvin}$, consistent with higher-lying CEF contributions.
Together, these observations indicate that the permittivity in \ce{PrMgAl11O19} reflects a superposition of low-energy quasi-doublet physics, rather than the simpler situation in \ce{CeMgAl11O19} where CEF excitations mediate the magnetoelectric response more indirectly.
In \ce{PrMgAl11O19}, the dielectric anomaly is therefore more naturally interpreted as arising from a direct coupling to the exchange-renormalized quasi-doublet sector, with overlapping features preventing the assignment of a unique anomaly temperature once $\mu_{0}H\gtrsim\SI{2}{\tesla}$.

\subsection{Thermal expansion}

\begin{figure*}[t]
\centering
\includegraphics[width=0.48\textwidth]{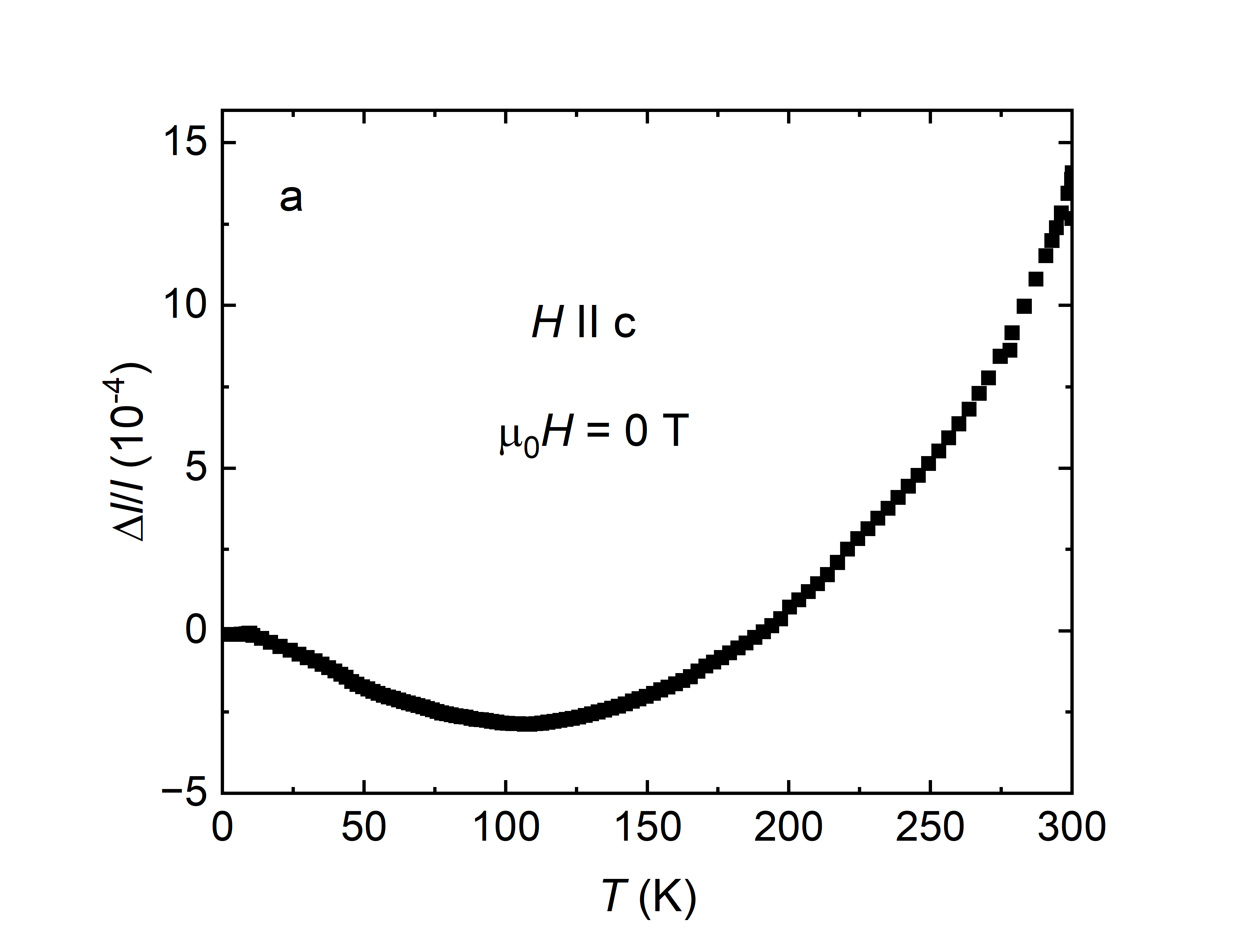}\hfill
\includegraphics[width=0.48\textwidth]{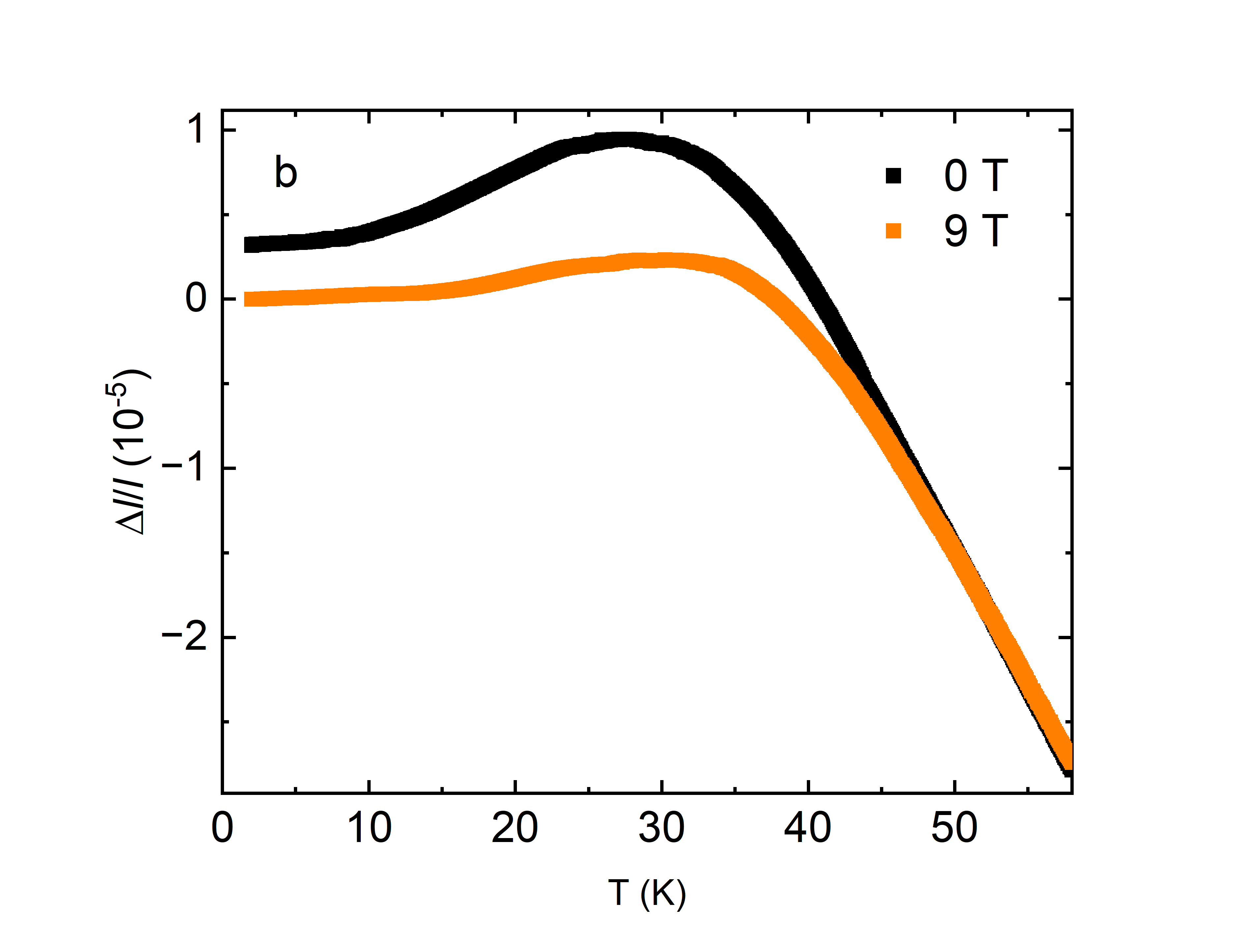}
\caption{(a) Uniaxial thermal expansion $\Delta l/l$ of \ce{PrMgAl11O19} measured along the $c$ axis in zero magnetic field, shown up to 300\,K.
(b) Low-temperature thermal expansion $\Delta l/l$ along the $c$ axis measured in zero field and in $\mu_{0}H=\SI{9}{\tesla}$ applied along $c$, highlighting the field evolution of the low-$T$ anomalies.}
\label{fig:Fig2}
\end{figure*}

The uniaxial thermal expansion of \ce{PrMgAl11O19} shares important similarities with the isostructural dipolar reference compound \ce{EuAl12O19}~\cite{bastien2024frustrated} (see Fig.~\ref{fig:Fig2}a).
$\Delta l/l$ measured along the $c$ axis shows a negative thermal expansion below \SI{100}{\kelvin}, consistent with the dipolar--lattice scenario proposed for \ce{EuAl12O19}, namely the slowing down of Al(5) vibration in the \ce{AlO5} bipyramids and the gradual formation of short-range antipolar correlations~\cite{bastien2024frustrated,Mittal2018PhononsNTE,Ramirez2000Underconstraint,Zhao2024AntiferroelectricNTE}.

A noticeable difference emerges below $\sim\SI{30}{\kelvin}$, where \ce{PrMgAl11O19} develops an additional broad hump that is not present in the Eu analogue.
Importantly, this hump appears in the same temperature range where the dielectric response deviates from the Barrett background and develops a pronounced low-$T$ feature. The near-saturation of $\Delta l/l$ below $\sim\SI{10}{\kelvin}$ is consistent with a regime where the response is dominated by the ground-state manifold~\cite{Kumar2025_PRB}.

Applying a magnetic field suppresses the low-temperature hump in $\Delta l/l$ and restores an \ce{EuAl12O19}-like, nearly temperature-independent behavior already below $\sim\SI{30}{\kelvin}$ by $\mu_0H=\SI{9}{\tesla}$. Since the permittivity anomaly in the same temperature window is suppressed and reshaped by field in a similar way, the thermal-expansion response is consistent with a coupled lattice--dielectric response at low temperatures. In this sense, the common temperature scale and field suppression provide an additional indication of magnetoelectric coupling, although thermal expansion alone cannot uniquely separate magnetoelectric effects from a more general magnetoelastic response~\cite{Ramirez2000Underconstraint,Wang2014PRX_AFE}.

\subsection{Microscopic interpretation: quasi-doublet physics and induced quantum magnetism}

At the microscopic level, the Pr quasi-doublet ($\Delta \approx 1.26$\,meV) in \ce{PrMgAl11O19} can be modeled as a two-level system in a transverse-field Ising (TFIM) framework~\cite{Chen2019_TFIM,Pfeuty1970_TFIM,Moessner2001_QFrust,Thalmeier2024_CEF}, similar to other Pr- and Tm-based non-Kramers quantum magnets such as \ce{Pr3BWO9}, \ce{PrTiNbO6}, and \ce{TmMgGaO4}~\cite{Li2018_PrTiNbO6,Li2020_TmMgGaO4,Liu2020_TmMgGaO4_Ising,Nagl2024_Pr3BWO9}. In this pseudospin-1/2 description, $\sigma_z$ represents the induced longitudinal moment and $\sigma_x$ mixes the two singlets. The ground-state physics of \ce{PrMgAl11O19} was previously shown to be well described by such a TFIM model~\cite{Kumar2025_PRB,Thalmeier2024_CEF}, whose single-ion part takes the form
\begin{equation}
\mathcal{H}_{2\mathrm{L}}^{(0)}
= -\frac{\Delta}{2}\,\sigma_x - A H\,\sigma_z,
\qquad
A = \tfrac{1}{2} g_c \mu_B .
\label{eq:H2L0}
\end{equation}
While Eq.~\eqref{eq:H2L0} describes an isolated quasi-doublet, inter-site exchange in the TFIM framework renormalizes the
two-level thermodynamics by converting a single local gap into a weakly dispersive collective spectrum. As a result,
the Schottky-like specific-heat feature expected for independent two-level systems is broadened \cite{Zheng2023_KTmSe2} and can shift because
the relevant excitation scale is effectively replaced by a renormalized $\Delta_{\mathrm{eff}}(T,H)$ set by exchange and
field, rather than a strict constant $\Delta$.

In the presence of an external electric field $E$, the effective electric-dipole matrix element $P_{\mathrm{eff}}$ of the Pr quasi-doublet couples to the transverse channel and modifies the singlet--singlet splitting. Symmetry requires that the level spacing be an even function of $E$, which we capture by an $E$-dependent splitting
\begin{equation}
\Delta'(E) = \sqrt{\Delta^2 + 4P_{\mathrm{eff}}^{\,2}E^2}.
\end{equation}
Projecting the TFIM Hamiltonian onto the lowest quasi-doublet in the presence of both $H$ and $E$ then yields the effective two-level Hamiltonian used in our analysis,
\begin{equation}
\label{eq:H2L_main}
\mathcal{H}_{2\mathrm{L}}
= -\frac{\Delta'(E)}{2}\,\sigma_x - A H\,\sigma_z,
\end{equation}
which serves as the starting point for the microscopic derivation of the biquadratic magnetoelectric coupling \cite{SI}. Because the spectrum derived from Eq.~\eqref{eq:H2L_main} is even in both $E$ and $H$, no linear $EH$ term appears, and the leading magnetoelectric invariant at the macroscopic level is biquadratic, $\lambda P^2 M^2$, as required by inversion symmetry.

\begin{figure*}[t]
\centering
\includegraphics[width=0.48\textwidth]{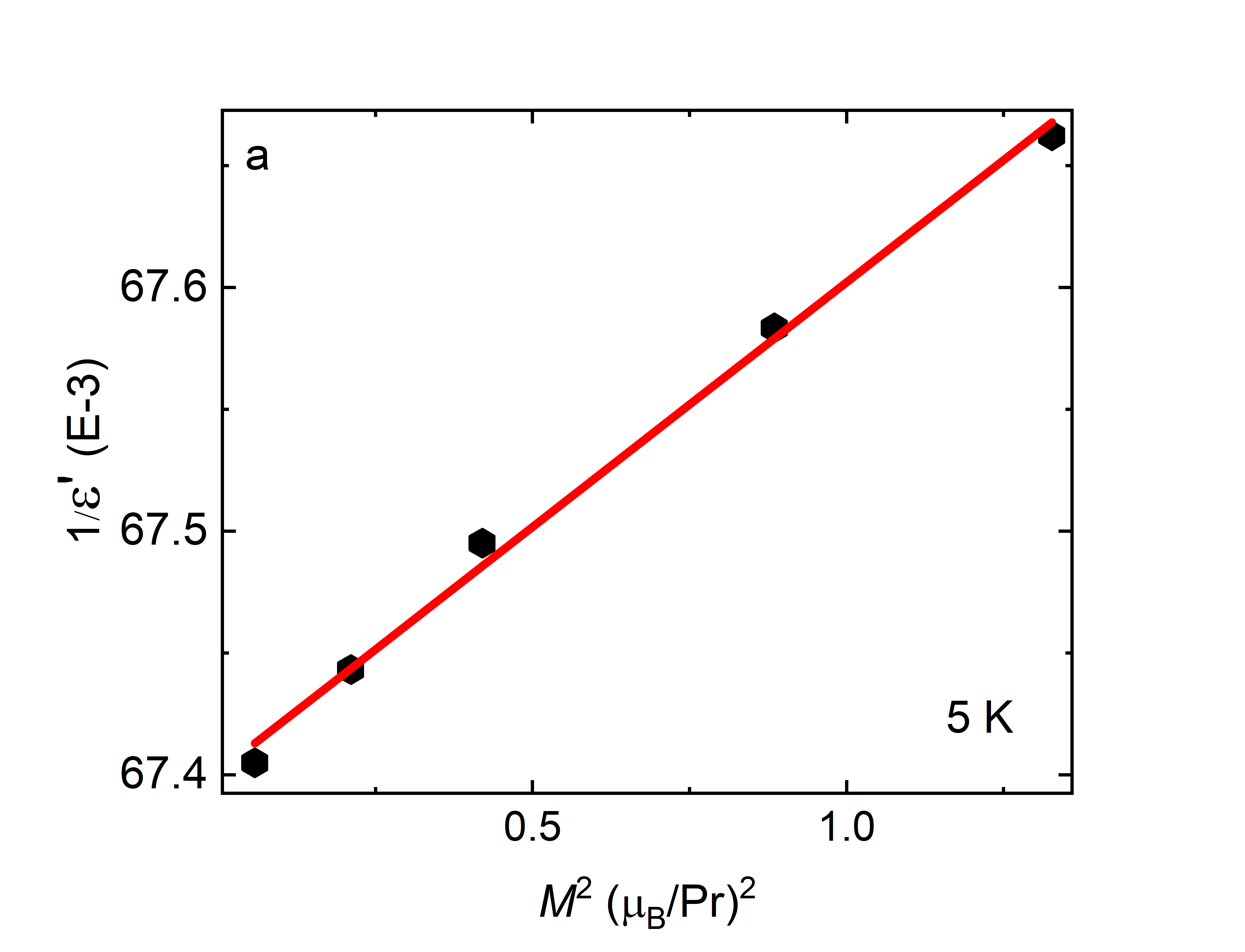}\hfill
\includegraphics[width=0.48\textwidth]{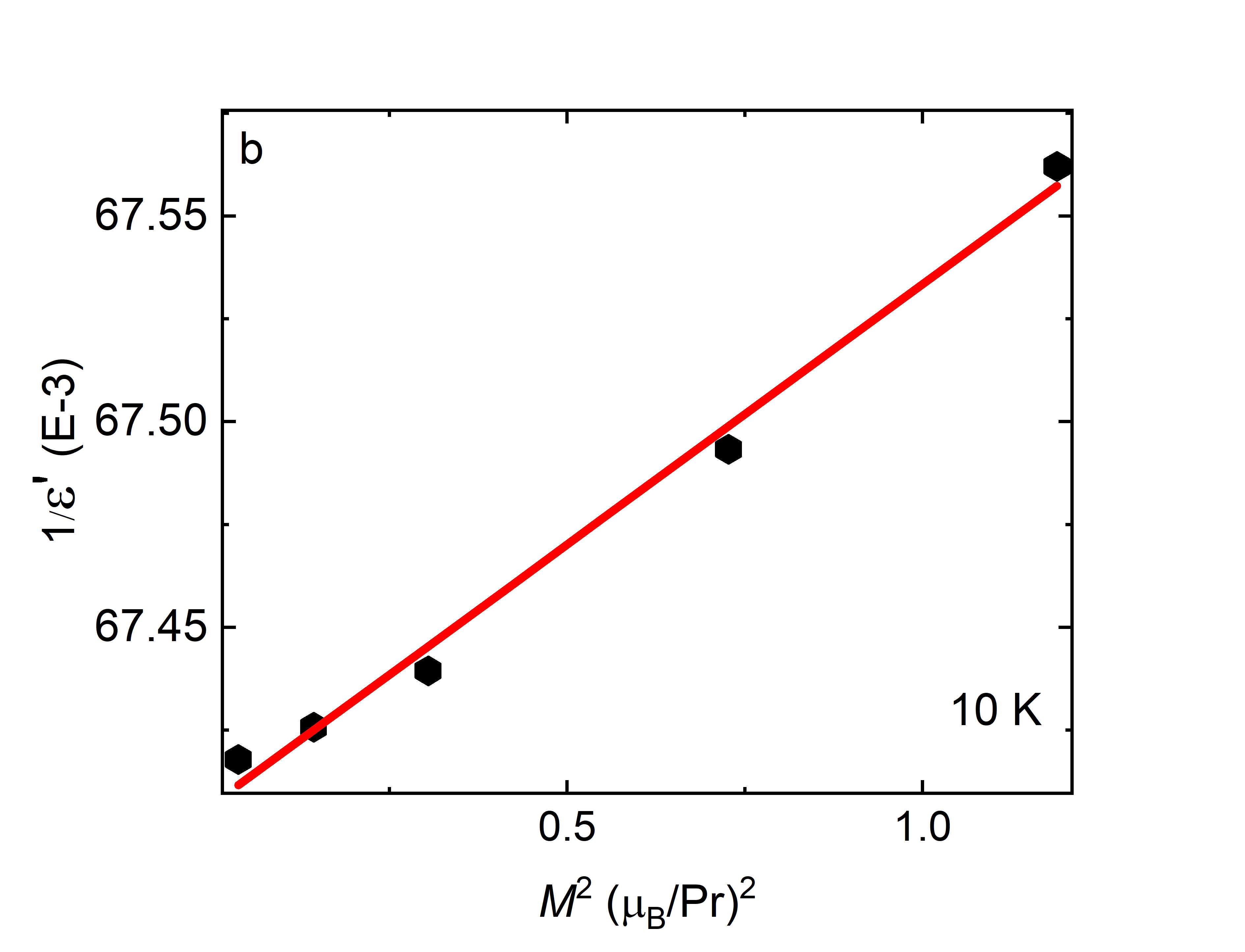}
\caption{(a) Inverse permittivity $\varepsilon'^{-1}$ versus $M^{2}$ at 5\,K. Solid line: linear fit yielding $\lambda_{5\,\mathrm{K}}$.
(b) Linear scaling of $\varepsilon'^{-1}$ with $M^{2}$ at 10\,K, giving $\lambda_{10\,\mathrm{K}}$.}
\label{fig:Fig3}
\end{figure*}

The detailed derivation of the magnetoelectric coupling from Eq.~\eqref{eq:H2L_main} is presented in the Supplementary Information \cite{SI}. In brief, relating $P_{\mathrm{eff}}$ to the dielectric oscillator strength $\Delta\epsilon_1$ of the Pr quasi-doublet via a Lorentz-oscillator estimate,
\begin{equation}
\Delta\epsilon_1 \simeq \frac{2 n_{\mathrm{Pr}} P_{\mathrm{eff}}^{\,2}}{\varepsilon_0 \Delta},
\end{equation}
and using $E = P/(\varepsilon_0 \varepsilon_r)$ and $H = M/\chi_c$ leads, up to numerical factors of order unity, to the scaling
\begin{equation}
\label{eq:lambda_scaling_main}
\lambda(T) \;\propto\;
\frac{(g_c \mu_B)^2}{\Delta^{2}}\,
\frac{\Delta \epsilon_1}{\varepsilon_0\, \varepsilon_r(T)^{2}\, \chi_c(T)^{2}}.
\end{equation}
Equation~\eqref{eq:lambda_scaling_main} shows that the biquadratic magnetoelectric coupling is governed by the CEF gap $\Delta$, the longitudinal $g$-factor $g_c$, the dielectric strength $\Delta\epsilon_1$ of the Pr quasi-doublet, and the $c$-axis susceptibility $\chi_c(T)$. While the two-level TFIM-based model captures the correct symmetry requirements and functional dependence of the coupling, a quantitative determination of $\lambda(T)$ requires precise knowledge of these microscopic parameters, including the effective dipole strength, higher CEF contributions, and the detailed temperature evolution of $\varepsilon_r(T)$ and $\chi_c(T)$. A comprehensive theoretical treatment together with dedicated experimental probes (such as a full CEF scheme determination) will therefore be necessary to fully resolve the microscopic origin and magnitude of the magnetoelectric coupling.

To nevertheless provide a direct quantitative check of the temperature dependence within the minimal two-level framework, we use the explicit SI result
$\lambda(T)=\Gamma_H(T)/\{[\varepsilon_0\varepsilon_r(T)]^2\chi_c(T)^2\}$ with
$\Gamma_H(T)\propto f(\Delta/k_BT)$ and
$f(x)=\tanh(x/2)-(x/2)\,\mathrm{sech}^2(x/2)$ \cite{SI}.
For $\Delta=1.26$\,meV this yields $f(\SI{5}{\kelvin})\approx 0.615$ and $f(\SI{10}{\kelvin})\approx 0.177$.
Experimentally, the zero-field permittivity is essentially unchanged between \SI{5}{\kelvin} and \SI{10}{\kelvin},
so $\varepsilon_r(T)$ cancels in the ratio. Using $\chi_c(\SI{5}{\kelvin})\approx 0.16$ and $\chi_c(\SI{10}{\kelvin})\approx 0.12$
(emu\,mol$^{-1}$\,Oe$^{-1}$), we obtain
\begin{equation}
\frac{\lambda(\SI{10}{\kelvin})}{\lambda(\SI{5}{\kelvin})}
=
\frac{f(\SI{10}{\kelvin})}{f(\SI{5}{\kelvin})}
\left[\frac{\chi_c(\SI{5}{\kelvin})}{\chi_c(\SI{10}{\kelvin})}\right]^2
\approx
\frac{0.177}{0.615}\left(\frac{0.16}{0.12}\right)^2
\approx 0.51.
\end{equation}
This is in good agreement with the Landau-fit result from Fig.~\ref{fig:Fig3},
$\lambda_{10\,\mathrm{K}}/\lambda_{5\,\mathrm{K}}\approx (4.77\times10^{-5})/(1.07\times10^{-4})\approx 0.45$.
The small deviation is plausibly due to exchange renormalization and other beyond-minimal effects (e.g., weak contributions from higher CEF levels), which are not captured by the strict two-level approximation.
This comparison confirms that, apart from an overall scale set by the (not independently determined) dipole strength
(or equivalently $P_{\mathrm{eff}}$), the minimal quasi-doublet model captures the observed weakening of
the biquadratic magnetoelectric coupling upon heating. At lower temperatures, where exchange-driven correlations become
more important, the same physics that renormalizes the Schottky anomaly can also renormalize $\lambda(T)$ by effectively
modifying the low-energy scale entering $f(\Delta/k_BT)$ and by changing $\chi_c(T)$ through collective (rather than purely
single-ion) response.

The splitting of the non-Kramers $J=4$ doublet arises from local symmetry lowering (e.g., Pr off-centering), consistent with time-reversal symmetry but unprotected by Kramers’ theorem \cite{Kumar2025_PRB}. This permits both a cubic (quadratic-in-$M$) coupling $\eta P M^{2}$ and the biquadratic term $\lambda P^{2} M^{2}$, although only the latter is allowed by the global inversion symmetry of the bulk. In contrast, \ce{CeMgAl11O19} retains a Kramers-protected ground doublet, which strictly forbids the cubic term and leaves $\lambda P^{2} M^{2}$ as the leading symmetry-allowed magnetoelectric invariant \cite{Kumar_CeMgAl11O19}.

In \ce{PrMgAl11O19}, the situation is complicated by the emergence of two overlapping low-temperature dielectric anomalies whose line shape closely resembles the double-hump structure observed in the magnetic specific heat \cite{Kumar2025_PRB}. This makes a reliable quantification of the cubic coupling $\eta$ impossible within the present dataset. Nevertheless, the temperature and field evolution of the dielectric response clearly indicates that the magnetoelectric coupling is driven by crystal-field physics of the Pr$^{3+}$ quasi-doublet---exactly as in \ce{CeMgAl11O19}---and is closely linked to the low-energy magnetic entropy and specific-heat features of the system.

\subsection{Comparison to Ce- and Eu-based hexaaluminates}

A comparative analysis of \ce{PrMgAl11O19} with \ce{CeMgAl11O19} and \ce{EuAl12O19} is instructive, as all three compounds belong to the hexaaluminate family and host frustrated \ce{AlO5} dipoles that underpin quantum paraelectric behavior. In \ce{EuAl12O19}, however, the Eu$^{2+}$ ion ($L=0$) lacks orbital degrees of freedom, so no spin-orbit-driven ME coupling emerges despite a dipolar-liquid-like state~\cite{bastien2024frustrated}. In \ce{EuAl12O19}, the permittivity is governed by local dipole disorder and remains decoupled from magnetism. In contrast, in \ce{CeMgAl11O19}, the dominant magnetoelectric coupling arises from virtual excitations to higher CEF doublets ($\Delta \sim 14\,\text{meV}$), with $\alpha_P$ renormalization producing a broad dielectric dip that persists up to \SI{5}{\tesla}~\cite{Cao2025}. In \ce{PrMgAl11O19}, the much lower-energy quasi-doublet splitting ($\Delta \sim 1.26\,\text{meV}$) permits both biquadratic and effective cubic channels, amplifying low-temperature responses. The three materials thus represent complementary limits of the same mechanism: Eu exemplifies dipole frustration without orbital activity, Ce embodies Kramers-protected high-energy excitations with stiffness renormalization, and Pr highlights the role of local symmetry breaking within a quasi-doublet. These differences underscore the essential role of orbital-active rare-earth ions in enabling ME interactions in hexaaluminates and the tunability of magnetoelectric responses in this family.

The absence of magnetic and dielectric long-range order down to 0.4~K and 0.3~K, respectively, together with field-tunable cubic and biquadratic invariants, suggests that \ce{PrMgAl11O19} provides a useful platform for exploring the interplay of quantum paraelectricity, ground-state CEF quasi-doublet splitting, and magnetic interactions. These results indicate that non-Kramers hexaaluminates may represent promising candidates for studying tunable magnetoelectric behavior in frustrated quantum paraelectrics.

\section{Conclusion}

\ce{PrMgAl11O19} shows characteristics of a non-Kramers quantum paraelectric with a frustrated antipolar lattice, where local symmetry lowering splits the Pr$^{3+}$ quasi-doublet and, in principle, allows both cubic and biquadratic magnetoelectric couplings. The observed linear scaling of $\varepsilon'^{-1}$ with $M^{2}$ is consistent with a quadratic channel. A second signature of magnetoelectric coupling is the low-temperature permittivity anomaly, whose field-dependent evolution closely mirrors the double-hump behavior observed in the magnetic specific heat, indicating a common origin in low-energy CEF-driven entropy redistribution and dipolar dynamics. Importantly, this work establishes the first experimental demonstration and quantitative Landau description of magnetoelectric coupling in the non-Kramers hexaaluminate \ce{PrMgAl11O19}, enabling a direct comparison to Kramers analogues within the same structural family. While the broad anomalies and multi-component magnetic response introduce uncertainty in the coupling extraction, \ce{PrMgAl11O19} provides a platform to investigate microscopic mechanisms of magnetoelectricity in frustrated rare-earth systems. Further high-resolution local probes will be valuable to clarify the relative roles of cubic and biquadratic terms in shaping the low-temperature dielectric response.

\section*{Acknowledgments}

We acknowledge funding from Charles University in Prague within the Primus research program with grant No. PRIMUS/22/SCI/016, the Grant Agency of Charles University (grant No. 438425) and the Czech Science Foundation (project No. 24/10791S). Crystal growth, structural analysis, and magnetic properties measurements were carried out in the MGML (\url{http://mgml.eu/}), supported within the Czech Research Infrastructures program (project no. LM2023065).
We thank R.~H.~Colman for supervision of the single-crystal growth of \ce{PrMgAl11O19} reported in our previous work~\cite{Kumar2025_PRB}.

\section*{Author Contributions}

S.K.\ carried out crystal growth and sample preparation, participated in all experiments, and performed all data analysis and interpretation. S.K.\ also wrote the manuscript.
G.B.\ initiated the project, contributed to the magnetization and dielectric measurements, and supervised the research.
P.P.\ and M.S.\ performed the dielectric measurements.
R.H.C.\ supervised crystal growth.
M.V.\ supervised the thermal expansion measurements and contributed to their analysis.
M.Ś.-B.\ arranged and supervised all work performed in Poznań.
M.K.\ and W.K.\ carried out the EPR measurements. S.K.\ supervised the dielectric experiments, contributed to data interpretation, and supervised the manuscript preparation.

\appendix
\begin{appendices}

\section*{S1. Microscopic Estimate of the Biquadratic Magnetoelectric Coupling}

The low-energy degrees of freedom of Pr$^{3+}$ in PrMgAl$_{11}$O$_{19}$ are governed by a non-Kramers quasi-doublet consisting of two singlets separated by $\Delta \approx 1.26$~meV. Following the standard pseudospin construction used for non-Kramers ions in transverse-field Ising-type models \cite{Pfeuty1970_TFIM,Chen2019_TFIM,Thalmeier2024_CEF,Moessner2001_QFrust}, the two-level system is described by Pauli matrices $\sigma_x$ and $\sigma_z$, where $\sigma_z$ corresponds to the induced longitudinal magnetic moment and $\sigma_x$ mixes the two singlets.

A magnetic field $H \parallel c$ couples to $\sigma_z$ with strength $A = \tfrac12 g_c \mu_B$. The electric field $E$ couples to the transverse dipole matrix element of the quasi-doublet, and the term $P_{\mathrm{eff}}E$ (with dimensions of energy) renormalizes the splitting between the singlets. The field-dependent splitting is
\begin{equation}
\Delta'(E) = \sqrt{\Delta^2 + 4P_{\mathrm{eff}}^{\,2}E^2}.
\end{equation}
The resulting two-level Hamiltonian is
\begin{equation}
\mathcal{H}_{2L} = -\frac{\Delta'(E)}{2}\,\sigma_x - A H\,\sigma_z .
\end{equation}
Diagonalization yields the two eigenenergies
\begin{equation}
\begin{aligned}
E_\pm(E,H) &= \pm \frac{1}{2} R(E,H), \\
R(E,H) &= \sqrt{\Delta^2 + 4P_{\mathrm{eff}}^{\,2}E^2 + 4A^2H^2}.
\end{aligned}
\end{equation}

Since $R(E,H)$ is even in both $E$ and $H$, no linear mixed term $EH$ can appear, consistent with the global inversion symmetry of the $P6_3/mmc$ structure.

At $T \to 0$, only the lower level contributes to the free energy,
\begin{equation*}
F_0(E,H) = -\frac{1}{2}R(E,H),
\end{equation*}
and expanding for small fields gives
\begin{equation*}
\begin{aligned}
F_0(E,H) \simeq{}& -\frac{\Delta}{2}
-\frac{P_{\mathrm{eff}}^{\,2}}{\Delta}E^2
-\frac{A^2}{\Delta}H^2 \\
&+\frac{P_{\mathrm{eff}}^{\,4}}{\Delta^3}E^4
+\frac{2A^2P_{\mathrm{eff}}^{\,2}}{\Delta^3}E^2H^2
+\frac{A^4}{\Delta^3}H^4.
\end{aligned}
\end{equation*}
Multiplying by the Pr density $n_{\mathrm{Pr}}$ gives a magnetoelectric cross term
\begin{equation*}
F \supset \Gamma_H(0)\,E^2H^2, \qquad
\Gamma_H(0)=2n_{\mathrm{Pr}}\frac{A^2P_{\mathrm{eff}}^{\,2}}{\Delta^3}.
\end{equation*}

At finite temperature, both levels contribute through the partition function
\begin{equation*}
Z(E,H)=2\cosh\!\left(\frac{\beta R(E,H)}{2}\right),
\qquad \beta = \frac{1}{k_B T}.
\end{equation*}
The free energy per Pr ion is therefore
\begin{equation*}
F(T,E,H) = -k_B T \ln\!\left[ 2\cosh\!\left(\frac{\beta R(E,H)}{2}\right)\right].
\end{equation*}
Expanding for small fields, the derivatives
\begin{equation*}
F'(R) = -\frac12 \tanh\!\left(\frac{\beta R}{2}\right), \qquad
F''(R) = -\frac{\beta}{4}\,\mathrm{sech}^2\!\left(\frac{\beta R}{2}\right)
\end{equation*}
evaluated at $R=\Delta$ give
\begin{equation*}
\Gamma_0(T) =
\frac{2A^2P_{\mathrm{eff}}^{\,2}}{\Delta^3}
\left[
\tanh\!\left(\frac{\Delta}{2k_BT}\right)
- \frac{\Delta}{2k_BT}\,\mathrm{sech}^2\!\left(\frac{\Delta}{2k_BT}\right)
\right].
\end{equation*}
Multiplying by $n_{\mathrm{Pr}}$ yields the free-energy density contribution

\begin{equation*}
\begin{aligned}
F(T,E,H) &\supset \Gamma_H(T)\,E^2H^2, \\
\Gamma_H(T) &=
2n_{\mathrm{Pr}}\frac{A^2P_{\mathrm{eff}}^{\,2}}{\Delta^3}\,
f\!\left(\frac{\Delta}{k_BT}\right).
\end{aligned}
\end{equation*}
where
\begin{equation*}
f(x) = \tanh\!\left(\frac{x}{2}\right)
- \frac{x}{2}\,\mathrm{sech}^2\!\left(\frac{x}{2}\right).
\end{equation*}

In the paraelectric regime the fields satisfy $E=P/(\varepsilon_0\varepsilon_r)$ and $H=M/\chi_c$, giving a Landau-level magnetoelectric term
\begin{equation}
F(P,M) \supset \lambda(T)\, P^2 M^2,
\qquad
\lambda(T)=
\frac{\Gamma_H(T)}{[\varepsilon_0\varepsilon_r(T)]^2 \chi_c(T)^2}.
\end{equation}

The expression above captures the symmetry-allowed biquadratic coupling and reproduces the experimentally observed scaling of $\varepsilon'^{-1}$ with $M^2$. Quantitative evaluation of $\lambda(T)$, however, depends sensitively on the dielectric oscillator strength of the Pr quasi-doublet and on additional effects such as higher-lying crystal-field levels, exchange interactions, and defect-induced dipolar dynamics. These contributions are not included in the minimal two-level model and may be required for complete quantitative agreement. Nevertheless, the derivation presented here establishes the microscopic origin and temperature dependence of the biquadratic magnetoelectric coupling in PrMgAl$_{11}$O$_{19}$.

\begin{figure*}[t]
  \centering
  \includegraphics[width=0.48\linewidth]{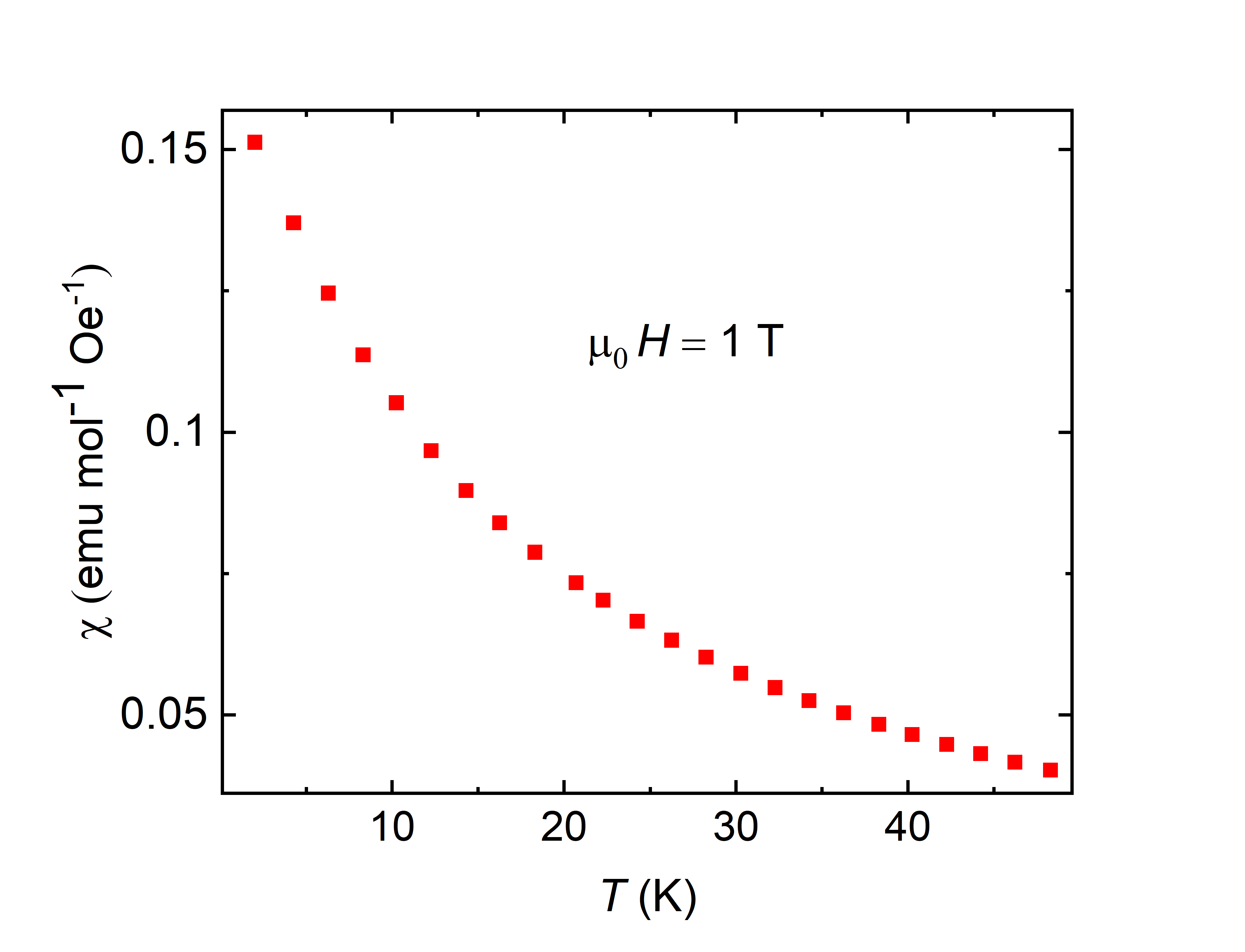}\hfill
  \includegraphics[width=0.48\linewidth]{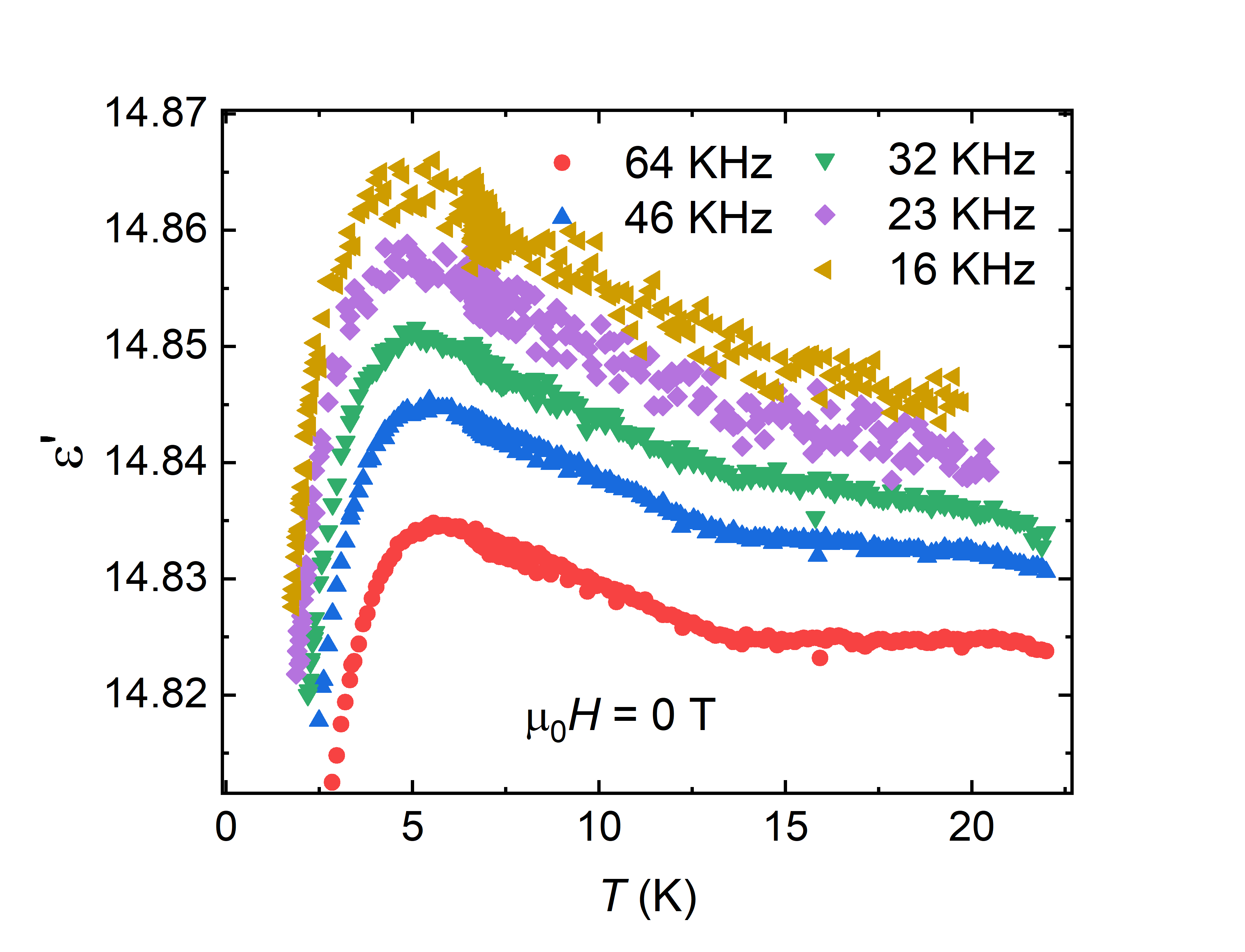}\\[0.5em]
  \includegraphics[width=0.48\linewidth]{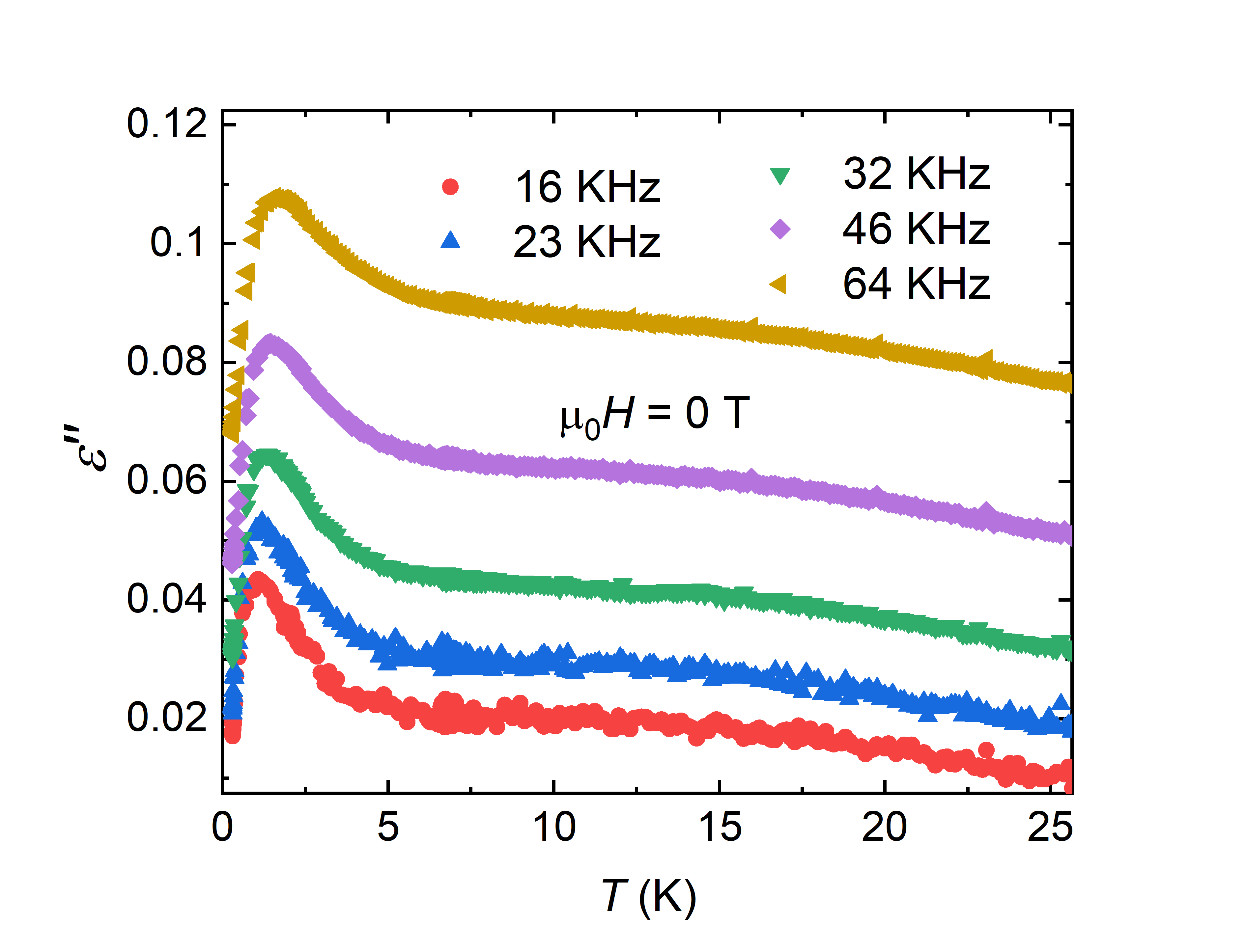}\hfill
  \includegraphics[width=0.48\linewidth]{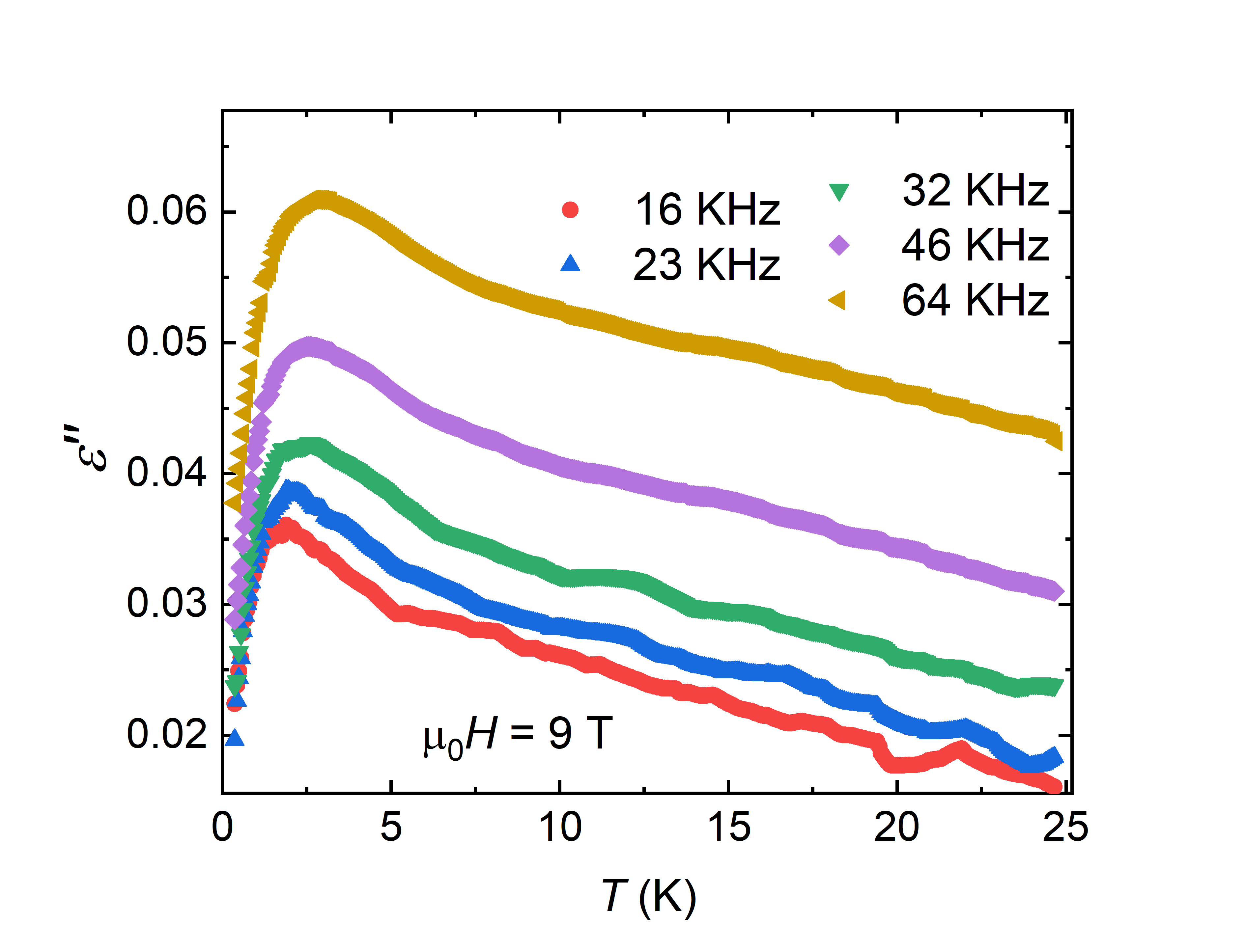}
  \caption{From top to bottom, left to right:
  (1) Magnetic susceptibility measured in an external field of 1~T, taken from Ref.~\cite{Kumar2025_PRB};
  (2) Temperature dependence of the real part of the permittivity $\varepsilon'(T)$
  at zero field for probe frequencies between 16 and 64~kHz;
  (3) Temperature dependence of the dielectric loss $\varepsilon''(T)$ under the same
  conditions;
  (4) Dielectric loss $\varepsilon''(T)$ at $\mu_{0}H = 9$~T for probe frequencies
  between 6 and 64~kHz.
  All measurements were performed with the electric field and/or magnetic field applied parallel to the crystallographic $c$ axis.}
  \label{fig:FigS}
\end{figure*}

\section*{S2. Frequency-Dependent Dielectric Response}

The frequency- and temperature-dependent dielectric response is summarized in Fig.~\ref{fig:FigS} (top right and bottom panels). Although measurements were performed from 1~Hz to 1~MHz, reliable data were obtained only in the range 16--64~kHz. At other frequencies the signal-to-noise ratio was significantly worse, which we attribute to the use of long coaxial cables in the cryostat setup and to the relatively small permittivity of the sample. The available frequency span is therefore insufficient for a reliable determination of a relaxation frequency from fits.

As illustrated in Fig.~\ref{fig:FigS}, at all temperatures examined we observe a slight decrease in $\varepsilon'(T)$ and an increase in $\varepsilon''(T)$ with increasing frequency, indicating that the relevant relaxation frequency lies above 64~kHz. Under an applied field of $\mu_{0}H = 9$~T, $\varepsilon''(T)$ retains a similar dispersive character across the same frequency window, with the anomaly modified in both amplitude and position compared to zero field.

\section*{S3. Angle-Dependent EPR Spectroscopy}

\begin{figure*}[t]
  \centering
  \includegraphics[width=0.48\linewidth]{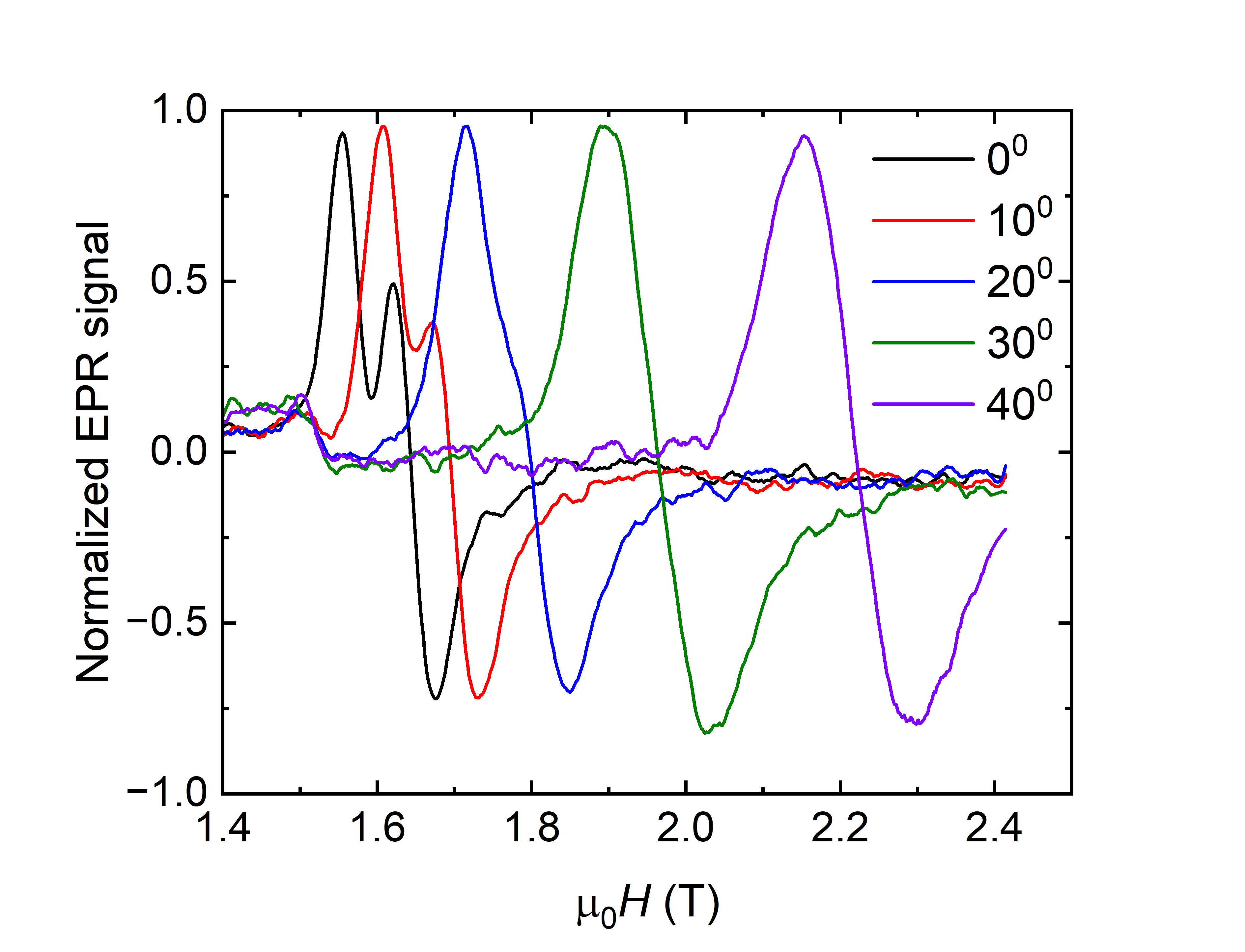}\hfill
  \includegraphics[width=0.48\linewidth]{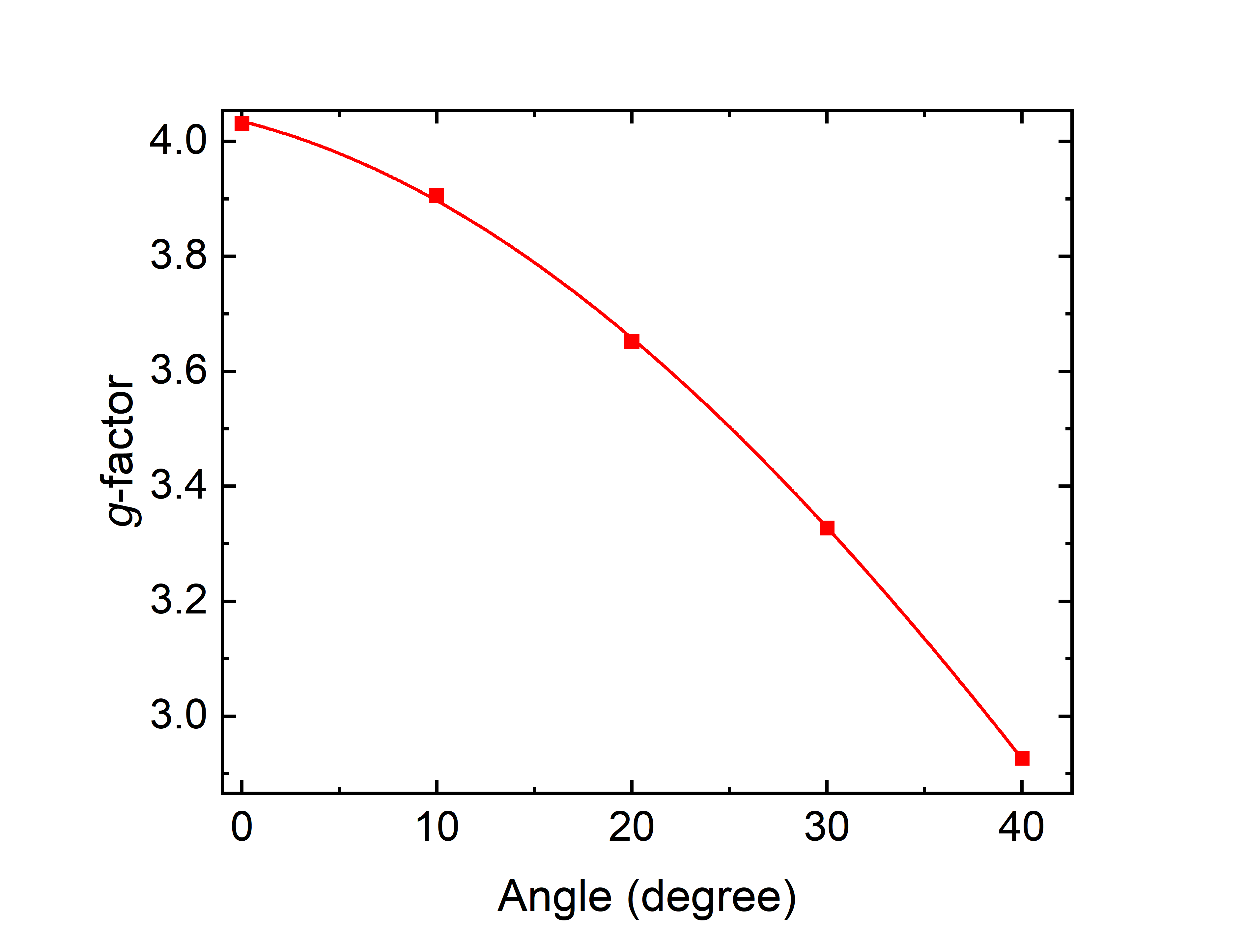}
  \caption{Left: Angle-dependent X-band EPR spectra of \ce{PrMgAl11O19} measured at $T=1.5$~K,
  plotted as the normalized first-derivative signal $dI/dB$ versus magnetic field for $\theta=0^\circ$--$40^\circ$
  (field rotated in a plane containing the crystallographic $c$ axis).
  Right: Effective $g$ factor obtained from the resonance field $B_{\rm res}$ extracted from the spectra as
  $g=h\nu/(\mu_{\rm B}B_{\rm res})$, plotted as a function of rotation angle $\theta$.
  The solid line shows a fit to the axial model.}
  \label{fig:SI_EPR}
\end{figure*}

Angle-dependent continuous-wave (CW) X-band electron paramagnetic resonance (EPR) measurements were performed on a single crystal of PrMgAl$_{11}$O$_{19}$ at $T=1.5$~K ($\nu \approx 9.4$~GHz) (Fig.~\ref{fig:SI_EPR}) using a goniometer, with the magnetic field rotated in a plane containing the crystallographic $c$ axis, following standard procedures for single-crystal EPR spectroscopy \cite{Bodziony2008JAC}. The rotation angle $\theta$ was defined such that $\theta=0^\circ$ corresponds to $H\parallel c$ (easy axis), whereas $\theta=90^\circ$ corresponds to $H\parallel ab$ (hard plane). Spectra were acquired in first-derivative mode, $dI/dB$, as a function of the magnetic field $B$. A linear background was subtracted from each spectrum; for visualization, the curves were normalized to unit amplitude (and lightly smoothed), while all parameter extraction was performed on the unsmoothed data.

In PrMgAl$_{11}$O$_{19}$, the EPR signal is weak and only partially resolved over a broad angular range. Consequently, we restrict the quantitative analysis to $\theta=0^\circ$--$40^\circ$, where a reproducible resonance feature is detectable. The resonance field was estimated using the extrema-midpoint procedure,
\begin{equation}
B_{\rm res}=\frac{B_{\max}+B_{\min}}{2},
\label{eq:Bres_midpoint}
\end{equation}
where $B_{\max}$ and $B_{\min}$ are the field positions of the positive and negative derivative extrema, respectively \cite{AbragamBleaney1986,Weil2006}. The effective $g$ factor was then obtained from
\begin{equation}
g=\frac{h\nu}{\mu_{\rm B}B_{\rm res}}.
\label{eq:g_from_Bres}
\end{equation}

The extracted $g(\theta)$ decreases rapidly as the field is tilted away from the $c$ axis, indicating pronounced uniaxial anisotropy (Fig.~\ref{fig:SI_EPR}). The angular evolution was analyzed within an axial $g$-tensor framework \cite{AbragamBleaney1986,Bodziony2008JAC}. Specifically, the data were fit using
\begin{equation}
g(\theta)=\sqrt{
g_c^2\cos^2\!\left(\theta-\theta_0\right)+
g_{ab}^2\sin^2\!\left(\theta-\theta_0\right)},
\label{eq:axial_g}
\end{equation}
yielding $g_c=4.07(1)$, $g_{ab}=1.42(3)$, and a small angular offset $\theta_0=-7.86(9)^\circ$ (accounting for mounting and alignment). Due to the weak and partially unresolved EPR response, these values should be regarded as effective parameters describing the angular trend rather than a high-precision determination of the $g$-tensor components. The robust outcome is the strong anisotropy $g_c \gg g_{ab}$, consistent with an essentially Ising-like response.

This qualitative EPR anisotropy agrees with our magnetization measurements (strongly suppressed in-plane response) and with previous bulk thermodynamic analyses that reported a dominant easy-axis $g$ factor together with Ising anisotropy \cite{Kumar2025_PRB}. We therefore conclude that, despite the limited EPR signal quality, the present angle-dependent spectra provide independent spectroscopic support for pronounced uniaxial anisotropy consistent with magnetization.

\section*{S4. Magnetoelastic versus magnetoelectric contributions to the thermal-expansion anomaly}

The broad hump observed in $\Delta l/l$ at low temperature and its suppression by magnetic field indicate that the length change is coupled to the same low-energy manifold that controls the magnetic and dielectric anomalies. At the phenomenological level, the leading couplings to a uniaxial strain $u\equiv \Delta l/l0$ can be written as a standard Landau expansion,
\begin{equation}
F = \frac{1}{2}C\,u^2 + g_M\,u\,M^2 + g_P\,u\,P^2 + \lambda\,P^2M^2 + \cdots ,
\label{eq:F_strain_couplings}
\end{equation}
where $C$ is the relevant elastic modulus, $g_M$ describes the direct magnetoelastic (magnetostriction-type) coupling, and $g_P$ describes electrostrictive coupling to the dielectric subsystem. Minimizing Eq.~\eqref{eq:F_strain_couplings} with respect to $u$ gives
\begin{equation}
u = -\frac{1}{C}\Big[g_M M^2 + g_P P^2 \Big] + \cdots .
\label{eq:u_basic}
\end{equation}
In zero applied electric field, $\langle P\rangle = 0$ but the dielectric subsystem contributes through fluctuations (or an effective $P^2$ scale), whose magnetic-field dependence arises because the ME term $\lambda P^2M^2$ renormalizes the quadratic dielectric stiffness. In the linear dielectric regime, this renormalization is directly measured via the Landau scaling $\varepsilon'^{-1}(T,H)=\alpha_P(T)+2\lambda(T)M^2(T,H)$, so that the fractional change of the dielectric stiffness is
\begin{equation}
\frac{\delta \alpha_P}{\alpha_P} \equiv
\frac{\varepsilon'^{-1}(T,H)-\varepsilon'^{-1}(T,0)}{\varepsilon'^{-1}(T,0)}
\simeq
\frac{2\lambda(T)M^2(T,H)}{\alpha_P(T)} .
\label{eq:frac_alphaP}
\end{equation}
Using the experimentally extracted $\lambda(T)$ and the measured magnetization, this fractional renormalization is small (typically $\delta \alpha_P/\alpha_P \ll 1$ in the field and temperature window of Fig.~3 in the main text), implying that the ME-induced correction to the strain via the dielectric channel in Eq.~\eqref{eq:u_basic} is expected to be a secondary effect. Therefore, the dominant contribution to the thermal-expansion hump is most naturally attributed to the direct magnetoelastic coupling $g_M u M^2$, while the magnetoelectric interaction provides a smaller renormalization consistent with the correlated anomalies observed in $\varepsilon'(T,H)$ and in the thermodynamic response. A quantitative separation of $g_M$ and the ME-induced correction would require additional inputs such as elastic constants and/or direct magnetostriction data, which are beyond the scope of the present work.
\end{appendices}

\bibliography{main}
\end{document}